\documentclass{emulateapj}
\lefthead{Dewangan, Griffiths \& Rao}

\slugcomment{Submitted to ApJ Letters}

\newcommand\chandra{{\it Chandra}}

\newcommand\xmm{{\it XMM-Newton}}

\newcommand\s{{\rm~s}}
\newcommand\ks{{\rm~ks}}
\newcommand\mhz{{\rm~mHz}}

\newcommand\hz{{\rm~Hz}}
\newcommand\kev{{\rm~keV}}

\newcommand\kms{\ifmmode {\rm~km\ s}^{-1} \else ~km s$^{-1}$\fi}
\newcommand\Hunit{\ifmmode {\rm~km\ s}^{-1}\ {\rm Mpc}^{-1}
        \else ~km s$^{-1}$ Mpc$^{-1}$\fi}
\newcommand\ctssec{\ifmmode {\rm~count\ s}^{-1} \else ~count s$^{-1}$\fi}
\newcommand\ergsec{\ifmmode {\rm~erg\ s}^{-1} \else
        ~erg s$^{-1}$\fi}
\newcommand\funit{\ifmmode {\rm~erg\ s}^{-1}\;{\rm cm}^{-2} \else
        ~ergs s$^{-1}$ cm$^{-2}$\fi}
\newcommand\phflux{\ifmmode {\rm~photon\ s}^{-1}\;{\rm cm}^{-2}
        \else   ~photon s$^{-1}$ cm$^{-2}$\fi}
\newcommand\efluxA{\ifmmode {\rm~erg\ s}^{-1}\;{\rm cm}^{-2}\;{\rm
        \AA}^{-1} \else ~erg s$^{-1}$ cm$^{-2}$ \AA$^{-1}$\fi}
\newcommand\efluxHz{\ifmmode {\rm~erg\ s}^{-1}\;{\rm cm}^{-2}\;{\rm
        Hz}^{-1} \else ~erg s$^{-1}$ cm$^{-2}$ Hz$^{-1}$\fi}
\newcommand\cc{\ifmmode {\rm~cm}^{-3} \else cm$^{-3}$\fi}
\newcommand\FWHM{\ifmmode {\rm~FWHM} \else ${\rm~FWHM}$\fi}
\newcommand\Msun{\ifmmode M_{\odot} \else $M_{\odot}$\fi}
\newcommand\Lsun{\ifmmode L_{\odot} \else $L_{\odot}$\fi}
\newcommand\ltsim{\raisebox{-.5ex}{$\;\stackrel{<}{\sim}\;$}}
\newcommand\gtsim{\raisebox{-.5ex}{$\;\stackrel{>}{\sim}\;$}}
\newcommand\hbeta{\ifmmode {\rm H}\beta \else H$\beta$\fi}
\newcommand\Kalpha{\ifmmode {\rm K}\alpha \else K$\alpha$\fi}
\newcommand\nh{\ifmmode N_{\rm H} \else N$_{\rm H}$\fi}
\usepackage{graphicx}
\usepackage{here}

\begin{document}

\title{Quasi Periodic Oscillations and Strongly Comptonized X-ray emission from Holmberg IX X-1}

\author{G. C. Dewangan\altaffilmark{1}, R. E. Griffiths\altaffilmark{1} \& A. R. Rao\altaffilmark{2}}
\altaffiltext{1}{Department of Physics, Carnegie Mellon University,
  5000 Forbes Avenue, Pittsburgh, PA 15213 USA; {\tt email: gulabd@cmu.edu} }
\altaffiltext{2}{Department of Astronomy \& Astrophysics, Tata Institute of Fundamental Research, Mumbai, 400005 India; {\tt email: arrao@tifr.res.in}}

\begin{abstract}
  We report the discovery of a $200\mhz$ quasi-periodic oscillation
  (QPO) in the X-ray emission from a bright ultra-luminous X-ray
  source (ULX) Holmberg IX X-1 using a long \xmm{} observation.  The
  QPO has a centroid at $\nu_{QPO} = 202.5_{-3.8}^{+4.9}\mhz$, a
  coherence $Q \equiv \nu_{QPO}/\Delta\nu_{FWHM} \approx 9.3$ and an
  amplitude (rms) of $6\%$ in the $0.2-10\kev$ band. This is only the
  second detection of a QPO from an ULX, after M~82 X-1, and provides
  strong evidence against beaming. The power spectrum is well fitted
  by a power law with an index of $\approx 0.7$.  The total integrated
  power (rms) is $\approx 9.4\%$ in the $0.001-1\hz$ range. The X-ray
  spectrum shows clear evidence for a soft X-ray excess component that
  is well described by a multicolor disk blackbody ($kT_{in}\sim
  0.3\kev$) and a high energy curvature that can be modeled either by
  a cut-off power law ($\Gamma \sim 1$; $E_{cutoff} = 9\kev$) or as a
  strongly Comptonized continuum in an optically thick ($\tau \approx
  7.3 $) and cool ($kT_e \approx 3\kev$) plasma.  Both the presence of
  the QPO and the shape of the X-ray spectrum strongly suggest that
  the ULX is not in the high/soft or thermally dominated state. A
  truncated disk and inner optically thick corona may explain the
  observed X-ray spectrum and the presence of the QPO.

\end{abstract}


\section{Introduction}
Ultra-luminous X-ray sources (ULXs) are extra-nuclear point X-ray
sources with luminosities exceeding the Eddington limit for a
$10{\rm~M_{\odot}}$ black hole (BH).  The most popular model to
explain the high luminosities of ULX is the ``intermediate mass BH''
with mass $M_{BH} \simeq 10^2 - 10^4{\rm~M_{\odot}}$ (e.g., Colbert \&
Mushotzky 1999).  Other popular models include X-ray binaries (XRBs)
with anisotropic emission (King et al. 2001), beamed XRBs with
relativistic jets directly pointing towards us i. e., scaled down
versions of blazars (Mirabel \& Rodriguez 1999), and XRBs with
super-Eddington accretion rates (Begelman 2002).

There is now extensive study of X-ray spectra of ULXs with \xmm{} \&
\chandra{}. ULXs show a variety of spectral shapes: ($i$) simple power
law similar to the low/hard state of BH XRBs (Winter, Mushotzky \&
Reynolds 2005), ($ii$) disk blackbody ($kT\sim 0.1-0.4\kev$) plus
power law similar to the high/soft state of BH XRBs (see e.g., Feng \&
Kaaret 2005), ($iii$) strongly curved spectra at high energies
(Stobbart, Roberts \& Warwick 2004; Feng \& Kaaret 2005; Agrawal \&
Misra 2006; Dewangan, Griffiths \& Rao 2006a; Stobbart, Roberts \&
Wilms 2006) without a corresponding spectral state in BH XRBs.  These
spectra are physically well described by a cool disk and thermal
Comptonization in an optical thick corona.  The cool disk plus
power-law spectra of ULXs may not always correspond to the high/soft
state of BH XRBs (see Roberts et al. 2005; Dewangan et al.  2006a).
The detailed variability properties of ULXs are yet to be known. It is
important to study both the spectral and temporal characteristics of
ULXs to understand their nature. BH XRBs exhibit characteristic power
density spectra (PDS) depending on their X-ray spectral state
(McClintock \& Remillard 2003). Many BH XRBs show quasi-periodic
oscillations (QPOs) and breaks in their PDS that represent
characteristic timescales close to the BH.  These characteristic
frequencies scale with the mass of the BH.  Therefore, determination
of the shape of the PDS and the detection of QPOs is crucial to
understand the nature of ULXs.  To date, there is only one ULX, M~82
X-1, that is known to show QPOs (Strohmayer \& Mushotzky 2003; Fiorito
\& Titarchuk 2004; Dewangan, Titarchuk \& Griffiths 2006b; Mucciarelli
et al. 2006).

In this {\it Letter}, we report the discovery of a $200\mhz$ QPO from
the bright ULX Holmberg IX X-1 (hereafter Ho~IX X-1; also known as M81
X-9) based on a long \xmm{} observation. The ULX is located in the
dwarf irregular galaxy Holmberg IX, a companion to M~81. The position
of the ULX is $\sim 2\arcmin$ away from the optical center of the host
galaxy.  Ho~IX X-1 is a bright X-ray source ($L_X \gtsim
10^{40}{\rm~erg~s^{-1}}$) that is variable on time scales of weeks and
months (La Parola et al. 2001).  Miller, Fabian \& Miller (2004)
presented X-ray spectra of Ho~IX~X-1 based on two snapshot \xmm{}
observations. The spectra were modeled by a cool ($kT\sim 0.17 -
0.29$) disk emission and simple power-law components.  Miller et al.
(2004) suggested a BH mass of $10^3M_{\odot}$ by scaling the disk
temperatures measured in Ho~IX~X-1 to those usually observed in BH
XRBs in their high state. Here we present power and energy spectral
analysis and show evidence for a QPO and high energy spectral
curvature.

\section{Observation \& Data Reduction}
\xmm{} observed Holmberg IX on 2004 September 26 for $119\ks$ .
The EPIC PN and MOS cameras were
operated in the large window  and the full frame modes,
respectively, using the thin filter. We used  SAS version 6.5.0 and
the most recent calibration database to process and filter the event
data. Examination of the background rate above $10\kev$ showed that
the observation is completely swamped by the particle background after
an elapsed time of $105\ks$ and this latter period was therefore
excluded from the rest of the analysis.  For temporal analysis, we
used all the PN and MOS events with patterns $\le 12$ and a continuous
exposure of $103.7\ks$ during which both the PN and MOS cameras
operated simultaneously. For spectral analysis, we chose a count rate
cut-off criterion to exclude the high particle background and used
event patterns $0-4$ (singles and doubles) and FLAG=0.
We extracted
events from the PN and MOS cameras using a $45\arcsec$ circular region
centered at the peak position of the ULX ($\alpha = 09^h57^m54^s$,
$\delta=69^{\circ}03{\arcmin}46{\arcsec}$; La Parola et al. 2001).
We also extracted appropriate background events within nearby circular
regions free of sources.

\section{The power spectrum}
\begin{figure}
  \centering \includegraphics[width=8cm]{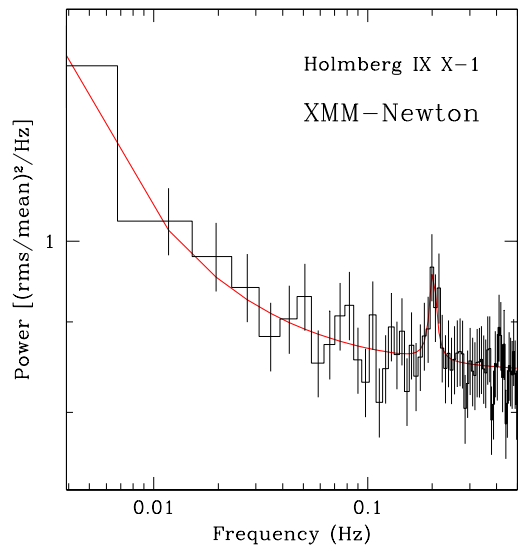}
  \caption{Power density spectrum of Holmberg IX X-1 derived from the
    EPIC PN and MOS data in the $0.2-10\kev$ band.  The white noise
    level expected ($\sim 0.8$) from the Poisson errors has not been
    subtracted.  The frequency resolution is $7.8{\rm~mHz}$. The
    best-fitting model comprising a simple power-law for the PDS
    continuum, a constant for the white noise and a Lorentzian for the
    QPO is shown as a thick line. }
  \label{f1}
\end{figure}

For temporal analysis, we combined the PN and MOS data to increase the
signal-to-noise ratio. We calculated a power density spectrum (PDS)
using the background corrected PN+MOS light curves sampled at $0.5\s$.
Figure~\ref{f1} shows the $0.2-10\kev$ PDS of Ho~IX~X-1 rebinned by a
factor of $1024$ yielding a frequency resolution of $7.8\mhz$. The PDS
continuum falls off approximately following a power law up to $\sim
60\mhz$ where the white noise arising from the Poisson errors starts
to dominate the continuum. The white noise continuum above $0.5\hz$ is
not shown in Fig.~\ref{f1}. There is a prominent QPO with its peak
frequency near $200\mhz$.

To fit the PDS, we used a simple power law for the continuum, a
Lorentzian to describe the QPO and a constant to account for the white
noise. This model resulted in an acceptable fit, providing a minimum
$\chi^2 = 28.4$ for $58$ degrees of freedom (dof). In Fig.~\ref{f1},
we show the best-fit model to the PDS as a thick line.  The errors on
the best-fit PDS model parameters, quoted below, are at a $1\sigma$
level.  The best-fit model resulted in a QPO centroid frequency
$\nu_{QPO} = 202.6\pm3.8\mhz$, a width $\nu_{FWHM} =
21.8_{-6.7}^{+10.5}\mhz$ and an amplitude $A_{QPO}=(3.7\pm0.9)\times
10^{-3}{\rm~(rms/mean)^2}$.  The best-fit power-law index and the
amplitude are $\Gamma = 0.73\pm0.02$ and $A_{PL} = (7.6\pm0.7)\times
10^{-3}{\rm~(rms/mean)^2/\hz}$ at $0.1\hz$, respectively. The best-fit
constant is $0.826\pm0.005{\rm~(rms/mean)^2/\hz}$.  The total
integrated power ($0.001-1\hz$) and the QPO power expressed as
$rms/mean$ are $9.4\%$ and $6\%$, respectively.  To estimate the
statistical significance of the detection of the QPO, we also fitted
the PDS with a simple power law and a constant. This model provided a
minimum $\chi^2 = 41.5$ for $63$ dof. Thus the addition of the
Lorentzian for the QPO was an improvement ($\Delta \chi^2 = - 13.1$
for three additional parameters) at a significance level of $99.5\%$
based on the maximum likelihood ratio test.

 \begin{figure}
   \centering \includegraphics[width=7cm,angle=-90]{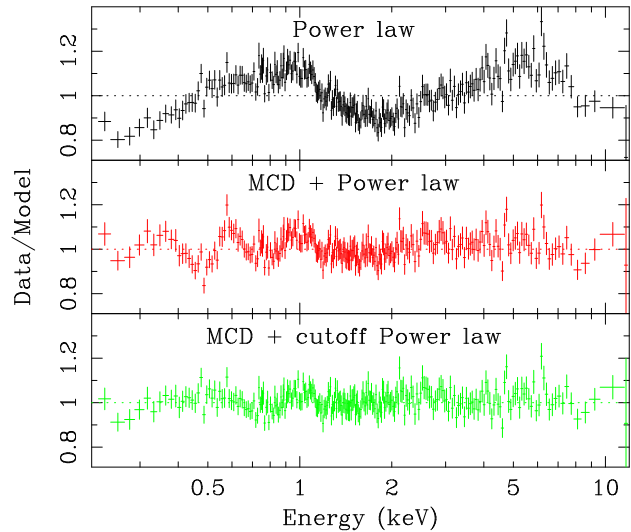}
   \caption{Ratios of EPIC PN data and the best-fit absorbed PL,
     MCD+PL and MCD+cutoff PL models. The soft X-ray excess and the
     high energy curvature are clearly seen in the top panel. The data have been rebinned for the purpose of display only.}
   \label{f2}
 \end{figure}

\begin{figure}
   \centering \includegraphics[width=7cm,angle=-90]{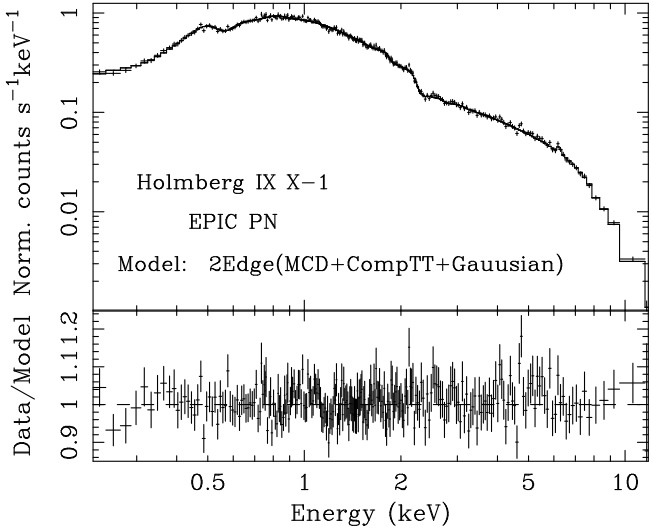}
   \caption{The EPIC PN data and the best-fit absorbed MCD {\it plus}
     thermal Comptonization model. Also shown is the ratio of the PN
     data and the best-fit model.}
   \label{f3}
 \end{figure}

\begin{deluxetable}{lcccc}
  \tabletypesize{\footnotesize}%
  \tablecolumns{3} \tablewidth{0pc} \tablecaption{Best-fit spectral
    model parameters for Holmberg IX X-1. \label{t1}} \tablehead{
  \colhead{Parameter} & \multicolumn{2}{c}{Model A\tablenotemark{a}} & \multicolumn{2}{c}{Model B\tablenotemark{a}} \\
  \colhead{} & \colhead{PN} & \colhead{MOS} & \colhead{PN} &
  \colhead{MOS} } \startdata
$N_H$\tablenotemark{b} & $1.85_{-0.07}^{+0.12}$  & $1.32_{-0.0.07}^{+0.10}$ & $1.78_{-0.12}^{+0.12}$ & $1.26_{-0.2}^{+0.2}$ \\
$E_{edge}$\tablenotemark{c} & $0.66_{-0.02}^{+0.02}$  & --  & $0.66_{-0.02}^{+0.02}$ & -- \\
$\tau$            & $0.16_{-0.04}^{+0.05}$  & -- & $0.16_{-0.02}^{+0.03}$  & -- \\
$E_{edge}$\tablenotemark{c}  & $7.89_{-0.40}^{+0.13}$  & -- & $7.88_{-0.42}^{+0.11}$  & -- \\
$\tau$            & $0.13_{-0.07}^{+0.08}$  & -- & $0.15_{-0.07}^{+0.09}$  & --  \\
$kT_{in}$\tablenotemark{c}  & $0.25_{-0.009}^{+0.007}$ & $0.36_{-0.03}^{+0.02}$ &  $0.28_{-0.02}^{+0.02}$ & $0.29_{-0.03}^{+0.04}$ \\
$A_{MCD}$\tablenotemark{d}  & $39.1_{-3.0}^{+4.2}$    & $6.3_{-0.9}^{+2.3}$ &  $30.8_{-8.9}^{+6.3}$ & $12.0_{-7.2}^{+6.5}$ \\
$\Gamma$                  & $1.02_{-0.15}^{+0.10}$ &  $0.35_{-0.22}^{+0.31}$  & -- & -- \\
$E_{cutoff}$ \tablenotemark{c}    & $9.0_{-1.8}^{+1.7}$    & $4.2_{-0.5}^{+1.6}$ & -- & --\\
$A_{PL}$\tablenotemark{d} &  $6.6_{-0.4}^{+0.5}$ & $4.5_{-0.5}^{+0.9}$ & -- & -- \\
$kT_{S}$\tablenotemark{c} & -- & -- & $0.51_{-0.06}^{+0.03}$ & $0.25_{-0.06}^{+0.05}$ \\
$kT_{e}$\tablenotemark{c}  & -- & -- & $3.0_{-0.36}^{+0.08}$ & $2.2_{-0.1}^{+0.2}$ \\
$\tau$ &  -- & -- & $7.3_{-0.5}^{+0.8}$ &  $9.9_{-0.9}^{+0.4}$  \\
$E_{K\alpha}$\tablenotemark{c}  &   $6.22_{-0.05}^{+0.03}$  &  -- &  $6.22_{-0.05}^{+0.05}$ &  --  \\
$\sigma$\tablenotemark{c}       &    $0.01$ (f) &  --&    $0.01$(f) &  --  \\
$f_{K\alpha}$\tablenotemark{e}         &   $1.4_{-0.8}^{+0.8}$     &  --  & $1.4_{-0.8}^{+0.8}$  &  --  \\
$EW_{K\alpha}$\tablenotemark{c} & $0.027$ & -- & $0.026$ & -- \\
$f_X$\tablenotemark{f}   & $6.0$    & $6.1$ & $6.0$ & $6.1$ \\
$L_X$\tablenotemark{f}   &  $8.3$       & $8.4$  & $8.3$  & $8.4$  \\
$\chi^2_{min}$ & 983.3 & $1965.0$ & $982.3$ & $1967.4$ \\
dof & 988 & 1903 & 987 & 1902 \\
\tableline
  \enddata
  \tablenotetext{a}{Model A: 2Edge(MCD+cutoff PL+Gaussian); Model B:
    2Edge(MCD+compTT+Gaussian). For the MOS data, the edges and the Gaussian line were
    excluded from the above models.}  \tablenotetext{b}{In units of
    $10^{21}{\rm~cm^{-2}}$.}  \tablenotetext{c}{In units of $\kev$.}
  \tablenotetext{d}{The MCD normalization
    $A_{MCD}=(R_{in}/km)/(D/10{\rm~kpc})$, where $R_{in}$ is the inner
    radius and $D$ is the distance. $A_{PL}$ is the power-law
    normalization in units of
    $10^{-4}{\rm~photons~cm^{-2}~s^{-1}~keV^{-1}}$ at $1\kev$.}
  \tablenotetext{e}{In units of
    $10^{-6}{\rm~photons~cm^{-2}~s^{-1}}$.}
  \tablenotetext{f}{Observed flux and luminosity in units of
    $10^{-12}{\rm~ergs~cm^{-2}~s^{-1}}$ and $10^{39}{\rm~erg~s^{-1}}$,
    respectively, and in the band of $0.2-10\kev$.}
    \end{deluxetable}

 \section{The energy spectrum}
 The PN and MOS spectra were grouped to a minimum
 of $50$ and $20$ counts per spectral channel, respectively, and
 analyzed with {\tt XSPEC 11.3}. The errors on the best-fit spectral
 parameters are quoted at a $90\%$ confidence level.
 First we fitted a simple absorbed power law (PL) model to the PN and
 MOS spectra of Ho~IX~X-1. We fit the MOS1 and MOS2 data jointly with
 an overall normalization constant to account for possible differences
 in the source extraction areas or calibration uncertainties. We used the
 $0.2-12\kev$ band in all the fits.  The simple power law model
 resulted in minimum $\chi^2 = 1850.3$ for $997$ dof and $2365.1$ for
 $1906$ dof for the PN and MOS data, respectively, thus providing
 statistically unacceptable fits to both the PN and MOS data. We have
 plotted the ratio of the PN data and the best-fit PL model in
 Figure~\ref{f2} ({\it top panel}). This plot clearly show a broad
 hump in the $3-8\kev$ band and a soft X-ray hump or excess emission
 below $1.5\kev$. The shape of the X-ray spectrum of Ho~IX~X-1 is very
 similar to that observed for NGC~1313 X-1 (Dewangan et al. 2006a).
 Similar ULX spectra have also been reported by Feng \& Kaaret (2005)
 and Stobbart et al. (2006). These spectra are well described by a
 steep power law at low energies and a hot accretion disk emission at
 high energies (Stoobart et al. 2006; Feng \& Kaaret 2005) or by a
 cool disk component and a cutoff power-law (Dewangan et al. 2006a).
 Addition of a hot multicolor disk blackbody (MCD) component improved
 the fit to the PN data significantly ($\chi^2 = 1149.5$ for 995 dof). 
 Significant improvement is found for the
 MOS data also.   Fig.~\ref{f2} ({\it middle panel}) shows the ratio
 of the PN data and the best-fit MCD+PL model. The high energy
 curvature is well described by the hot MCD component but the soft
 excess component is not well modeled by the steep power-law
 component.  Replacing the simple power-law with a cut-off power law
 improves the fit (PN: $\chi^2 = 1028.6$ for 994 dof and MOS: $\chi^2=
 1965.0$ for 1903 dof) and the MCD component results in cool disk
 ($kT_{in} \sim 0.25\kev$). Examination of the ratio of the PN data and
 the best-fit {\tt MCD+cutoff PL} model, shown in Fig.~\ref{f2} ({\it
   bottom panel}), reveals several weak features e.g., possible edges
 at $\sim 0.7$ and $\sim 8\kev$ and an emission feature at $\sim
 6\kev$. Addition of edges at $\sim 8\kev$ and $\sim 0.7\kev$ to the
 MCD+cutoff PL model improves the fits to the PN data by $\Delta \chi^2
 = -8.0$ and $-26.2$, respectively, for two additional parameters.
 Addition of a narrow Gaussian line at $\sim 6.2\kev$ improves the PN
 fit marginally ($\Delta \chi^2 = -11.0$ for two additional parameters).
 The best-fit parameters are listed in Table~\ref{t1}. The MOS data do
 not show evidence for any discrete feature, addition of edges or
 lines does not improve the fit.

 The high energy curvature or cut-off around $\sim 7\kev$ can
 physically be described as arising from thermal Comptonization in an
 optically thick corona (Goad et al. 2006; Agrawal \& Misra 2006;
 Stobbart et al. 2006). To test such a scenario for Ho~IX~X-1, we used
 the XSPEC model {\tt compTT} (Titarchuk 1994) that describes thermal
 Comptonization of soft photons in a hot plasma. The parameters of
 this model are the soft photon temperature ($kT_{S}$), electron
 plasma temperature ($kT_{e}$), plasma optical depth ($\tau$) and the
 normalization. We also used the MCD model to describe the soft excess
 component. The {\tt MCD+compTT} model provided minimum $\chi^2$
 values similar to that obtained for the {\tt MCD+cutoff PL} model.
 Additions of two absorption edges at $\sim 8\kev$ and $\sim 0.7\kev$
 and a narrow Gaussian line at $\sim 6.2\kev$ improved the fit to the
 PN data by $\Delta\chi^2 = -8.8$, $-30.1$ and $-8.2$, respectively.
 The best-fit parameters are listed in Table~\ref{t1}. The curvature
 or cut-off is well modeled by thermal Comptonization in an optically
 thick ($\tau = 7.3_{-0.5}^{+0.8}$ and relatively cool ($kT_e =
 3.0_{-0.4}^{+0.1}\kev$) electron plasma. Using the above model, we
 derive an unabsorbed flux of
 $1.06\times10^{-11}{\rm~ergs~cm^{-2}~s^{-1}}$ for Ho IX X-1 in the
 $0.1-100\kev$ band. This translates into a bolometric luminosity of
 $\sim 1.5\times 10^{40}{\rm~ergs~s^{-1}}$ using a distance of 3.4 Mpc
 (Georgiev et al. 1991; Hill et al. 1993).

\section{Discussion}
Based on a long \xmm{} observation, we have discovered a QPO with
centroid frequency $\nu_{QPO} = 202.5_{-3.8}^{+4.9}\mhz$ in the power
spectrum of Ho~IX~X-1.  The QPO has an amplitude (rms) of $6\%$ and a
coherence $Q\approx 9.3$.  The underlying PDS is well described by a
power law with an index of $\approx 0.7$.  This is only the second
detection of a QPO from any ULX.  M82 X-1 was the first ULX to show
clear evidence for a QPO (Strohmayer \& Mushotzky 2003).  M82 X-1
shows a variable QPO with the centroid frequency in the range of
$\approx 50 -170\mhz$ (Muccarelli et al. 2006). 
The $200\mhz$ QPO from Ho~IX~X-1 appears to be similar to the $54\mhz$
QPO (rms amplitude $\approx 8.5\%$; $Q\approx5$; Strohmayer \&
Mushotzky 2003), seen in the first \xmm{} observation of M82 X-1, in
terms of the amplitude and coherence.  The PDS continua of Ho IX X-1
and M~82 X-1 (in the first \xmm{} observation) are also simple power
laws without a clear evidence for a break.

BH XRBs show flat-topped, band-limited noise in their low state (LS).
The flat-topped noise weakens in the intermediate/very high states
(IS/VHS) and is absent in the high state (HS). Strong low frequency
QPOs are observed in the IS/VHS and occasionally in the LS but are
usually not present in the HS (see e.g., McClintock \& Remillard
2003). The IS/VHS are also known as the soft or hard intermediate
states (SIMS or HIMS), depending on whether the spectrum is soft or
hard, in the classification scheme of Belloni (2005). In the HIMS, BH
XRBs show strong $0.1-15\hz$ QPO with high Q-factor ($7-12$) and large
amplitude ($3-16\%$ rms). These are called type C QPOs (Casella et al.
2005).  The QPO in Ho~IX~X-1 (as well as in M82~X-1) have properties
similar to the type C QPOs and suggest that the variability properties
of these ULXs are likely to be similar to that of BH XRBs in their HIMS.

X-ray spectrum of Ho IX X-1 is curved at high energy ($E \gtsim
6\kev$) and shows a soft X-ray excess emission at low energies $\ltsim 1\kev$.
The PN spectrum is well described by an MCD ($kT \sim 0.25\kev$) and a
cut-off power law ($\Gamma \sim 1$; $E_{cutoff} \sim 9\kev$) or a {\tt
  compTT} component ($kT_S \sim 0.5\kev$; $kT_e \sim 3\kev$ and $\tau
\sim 7$). While this form of unusual X-ray spectrum is common among
bright ULXs (Stobbart et al. 2006), there is no corresponding spectral
states of BH XRBs. Both the presence of the QPO and the high energy
curvature in the X-ray emission from Ho~IX~X-1 are inconsistent with
the ULX being in the high/soft state and the soft X-ray excess
emission cannot be interpreted as the optically thick emission from a
disk extending down to the last stable orbit. Thus the
temperature of the soft excess component cannot be used to estimate
the BH mass for Ho IX X-1.

In BH XRBs, QPOs are observed when both the disk and power-law
components contribute significantly to the total X-ray emission (e.g.,
Sobczak et al. 2000). When strong QPOs are seen in the LS,
IS and VHS, the temperature of the disk
component is cooler than that in the HS. In these cases,
the inner disk is thought to be truncated (Done \& Kubota 2005).  The
presence of the strong QPO in the power spectrum of Ho IX X-1, then,
implies that its accretion disk must be truncated at some inner
radius, larger than the radius of the innermost stable orbit.  Done \&
Kubota (2005) have investigated accretion disk structure of BH XRBs in
their VHS characterized by strongly Comptonized spectra
and cooler disk temperatures compared to that in the HS.
They have proposed a model in which the corona and the disk are parts
of the same accretion flow. At high accretion rates, the inner corona
is energetically-coupled to an outer disk that is truncated at some
inner radius, thus resulting in a cooler disk and optically
thick inner corona. 
ULXs with cool soft X-ray emission and high energy curvature such as
Ho IX X-1 may be the extreme cases of VHS of BH XRBs.  In
Ho~IX~X-1, the disk truncation radius is likely to be larger and the inner
corona is optically thicker compared to that in the VHS of BH XRBs.
Thus Ho IX X-1 may be accreting at a rate (relative to the Eddington
rate) larger than that in the VHS of BH XRBs. Interestingly, some BH
XRBs e.g., GRS~1915+105 show spectral curvature around $30\kev$ in
their HIMS (Vadawale et al. 2001;  Kubota \&
Done 2004). Due to the presence of a strong QPO and spectral curvature
at $\sim 3\kev$ in ULXs, Ho~IX X-1 and M~82 X-1, similar to the type-C
QPO and high energy curvature at $\sim 30\kev$ in the HIMS of some BH
XRBs, it is quite conceivable that ULXs have their spectral curvature
at lower energies in their HIMS.

If the association of the observed QPO in ULXs to type-C QPO of BH
XRBs is correct, it gives a useful handle to constrain the mass of the
BH in Ho~IX~X-1. In the HIMS states, the accretion rate is typically
$>30\%$ of the Eddington luminsoity. Considering the fact that QPOs
are associated with accretion discs (and hence ruling out the
possibility of beamed jet emission for the observed X-ray emission),
this puts a strong upper limit of 400 M$_\odot$ for the mass of the BH
in Ho~IX~X-1. Further, the spectral curvatures in GRS 1915+105 are
preferentially seen just before radio flares and the QPO frequencies
during such states are in the range of $0.7-3\hz$ (Vadawale et al.
2003). If we equate this frequecy range to the observed QPO in Ho IX
X-1, we get an estimate for the BH mass in Ho IX X-1 as $50-200{\rm~M_{\odot}}$.

Associating the observed QPO in Ho~IX~X-1 to type C QPO in BH XRBs has
another interesting consequence. It is observed that transition from
HIMS to soft state (or SIMS) gives rise to fast relativistic jet in
galactic sources.  Similarly, such ultra-luminous jets should be seen
in ULXs, when they make a transition from HIMS to SIMS. Detecting such
relativistic jets in ULXs will put the association with the BH XRBs on
a firm footing.

\acknowledgements GCD acknowledges the support of NASA grants through
the awards NNG04GN69G and NNG05GN35G.  This work is based on
observations obtained with \xmm{}, an ESA science mission with
instruments and contributions directly funded by ESA Member States and
the USA (NASA).


\begin{thebibliography}{}

\bibitem[Agrawal \& Misra(2006)]{2006astro.ph..1325A} Agrawal, V.~K., \& 
Misra, R.\ 2006, \apjl, 638, 83 

\bibitem[Belloni(2005)]{2005astro.ph..7556B} Belloni, T.\ 2005, ArXiv 
Astrophysics e-prints, arXiv:astro-ph/0507556  

\bibitem[Casella et al.(2005)]{2005ApJ...629..403C} Casella, P., Belloni, 
T., \& Stella, L.\ 2005, \apj, 629, 403 
 

\bibitem[Colbert \& Mushotzky(1999)]{1999ApJ...519...89C} Colbert,
  E.~J.~M.~\& Mushotzky, R.~F.\ 1999, \apj, 519, 89 (CM99)

\bibitem[Dewangan et al.(2006a)]{2005astro.ph.11112D} Dewangan, G.~C., 
Griffiths, R.~E., \& Rao, A.~R.\ 2006a, submitted to ApJ, ArXiv Astrophysics e-prints, 
arXiv:astro-ph/0511112

\bibitem[Dewangan et al.(2006b)]{2006ApJ...637L..21D} Dewangan, G.~C., 
Titarchuk, L., \& Griffiths, R.~E.\ 2006b, \apjl, 637, L21 

 
\bibitem[Done \& Kubota(2005)]{2005astro.ph.11030D} Done, C., \& Kubota, 
A.\ 2005, ArXiv Astrophysics e-prints, arXiv:astro-ph/0511030 
 


\bibitem[Feng \& Kaaret(2005)]{2005ApJ...633.1052F} Feng, H., \& Kaaret, 
P.\ 2005, \apj, 633, 1052 

\bibitem[Fiorito \& Titarchuk(2004)]{2004ApJ...614L.113F} Fiorito, R.,
  \& Titarchuk, L.\ 2004, \apjl, 614, L113 


\bibitem[Georgiev et al.(1991)]{1991A&AS...89..529G} Georgiev, T.~B., 
Bilkina, B.~I., Tikhonov, N.~A., \& Karachentsev, I.~D.\ 1991, \aaps, 89, 
529 

\bibitem[Goad et al.(2006)]{2006MNRAS.365..191G} Goad, M.~R., Roberts, 
T.~P., Reeves, J.~N., \& Uttley, P.\ 2006, \mnras, 365, 191 

\bibitem[Hill et al.(1993)]{1993ApJ...402L..45H} Hill, J.~K., et al.\ 1993, 
\apjl, 402, L45 
  

\bibitem[King et al.(2001)]{2001ApJ...552L.109K} King, A.~R., Davies, 
M.~B., Ward, M.~J., Fabbiano, G., \& Elvis, M.\ 2001, \apjl, 552, L109 

\bibitem[Kubota \& Done(2004)]{2004MNRAS.353..980K} Kubota, A., \& Done, 
C.\ 2004, \mnras, 353, 980 


\bibitem[La Parola et al.(2001)]{2001ApJ...556...47L} La Parola, V., Peres, 
G., Fabbiano, G., Kim, D.~W., \& Bocchino, F.\ 2001, \apj, 556, 47 

\bibitem[McClintock \& Remillard(2003)]{2003astro.ph..6213M} McClintock, 
J.~E., \& Remillard, R.~A.\ 2003, ArXiv Astrophysics e-prints, 
arXiv:astro-ph/0306213 

\bibitem[Miller et al.(2004)]{2004ApJ...607..931M} Miller, J.~M., Fabian, 
A.~C., \& Miller, M.~C.\ 2004, \apj, 607, 931 
 
 
\bibitem[Mirabel \& Rodr{\'{\i}}guez(1999)]{1999ARA&A..37..409M} Mirabel, 
I.~F., \& Rodr{\'{\i}}guez, L.~F.\ 1999, \araa, 37, 409 

\bibitem[Mucciarelli et al.(2006)]{2006MNRAS.365.1123M} Mucciarelli, P., 
Casella, P., Belloni, T., Zampieri, L., \& Ranalli, P.\ 2006, \mnras, 365, 
1123 
 


\bibitem[Roberts et al.(2005)]{2005MNRAS.357.1363R} Roberts, T.~P.,
  Warwick, R.~S., Ward, M.~J., Goad, M.~R., \& Jenkins, L.~P.\ 2005,
  \mnras, 357, 1363

\bibitem[Soria et al.(2004)]{2004A&A...423..955S} Soria, R., Motch, C., 
Read, A.~M., \& Stevens, I.~R.\ 2004, \aap, 423, 955 

\bibitem[Stobbart et al.(2004)]{2004MNRAS.351.1063S} Stobbart, A.-M., 
Roberts, T.~P., \& Warwick, R.~S.\ 2004, \mnras, 351, 1063 

\bibitem[Stobbart et al.(2006)]{2006astro.ph..1651S} Stobbart, A., Roberts, 
T.~P., \& Wilms, J.\ 2006, ArXiv Astrophysics e-prints, 
arXiv:astro-ph/0601651 

\bibitem[Strohmayer \& Mushotzky(2003)]{2003ApJ...586L..61S}
  Strohmayer, T.~E., \& Mushotzky, R.~F.\ 2003, \apjl, 586, L61 (SM03)

\bibitem[Titarchuk(1994)]{1994ApJ...434..570T} Titarchuk, L.\ 1994, \apj, 
434, 570 

\bibitem[Titarchuk \& Fiorito(2004)]{2004ApJ...612..988T} Titarchuk,
  L., \& Fiorito, R.\ 2004, \apj, 612, 988 (TF04)

\bibitem[Vadawale et al.(2001)]{2001A&A...372..793V} Vadawale, S.~V., Rao, 
A.~R., \& Chakrabarti, S.~K.\ 2001, \aap, 372, 793 

\bibitem[Vadawale et al.(2003)]{}Vadawale, S. V., Rao, A. R., Naik,
  S., Yadav, J. S., Ishwara-Chandra, C. H., Pramesh Rao, A., Pooley,
  G. G. 2003, \apj, 597, 1023

 
\bibitem[Winter et al.(2005)]{2005astro.ph.12480W} Winter, L.~M., 
Mushotzky, R.~F., \& Reynolds, C.~S.\ 2005, ArXiv Astrophysics e-prints, 
arXiv:astro-ph/0512480 



\bibitem[Sobczak et al.(2000)]{2000ApJ...531..537S} Sobczak, G.~J.,
  et al., \ 2000a, \apj, 531, 537 

\end{thebibliography}
\end{document}